\documentclass[aip,reprint,amsmath,amssymb]{revtex4-1}

\usepackage{graphicx}

\newcommand{\V}[1]{\mathbf{#1}} 
 
\newcommand\Alfven{Alfv\'en }
\newcommand\Alfvenic{Alfv\'enic }
\newcommand{\figref}[1]{Figure~\ref{#1}}
\newcommand{\secref}[1]{\S\ref{#1}}
\newcommand{\xhat}{\mbox{$\hat{\mathbf{x}}$}} 
\newcommand{\yhat}{\mbox{$\hat{\mathbf{y}}$}} 
\newcommand{\zhat}{\mbox{$\hat{\mathbf{z}}$}}

\usepackage{color}

\begin{document}

\title{\Alfven Wave Collisions, The Fundamental 
Building Block of Plasma Turbulence III:
 Theory for Experimental Design} 

\author{G.~G. Howes}
\email[]{gregory-howes@uiowa.edu}
\affiliation{Department of Physics and Astronomy, University of Iowa, Iowa City, 
Iowa 52242, USA}
\author{K.~D.~Nielson}
\affiliation{Department of Physics and Astronomy, University of Iowa, Iowa City, 
Iowa 52242, USA}
\author{D.~J.~Drake}
\affiliation{Department of Physics, Astronomy, and Geosciences, Valdosta State University, Valdosta, Georgia 31698, USA}
\author{J.~W.~R.~Schroeder}
\affiliation{Department of Physics and Astronomy, University of Iowa, Iowa City, 
Iowa 52242, USA}
\author{F.~Skiff}
\affiliation{Department of Physics and Astronomy, University of Iowa, Iowa City, 
Iowa 52242, USA}
\author{C.~A.~Kletzing}
\affiliation{Department of Physics and Astronomy, University of Iowa, Iowa City, 
Iowa 52242, USA}
\author{T.~A.~Carter}
\affiliation{Department of Physics and Astronomy, University of California, Los Angeles, California 90095-1547, USA}

\date{\today}

\begin{abstract}
Turbulence in space and astrophysical plasmas is governed by the
nonlinear interactions between counterpropagating \Alfven waves.  Here
we present the theoretical considerations behind the design of the
first laboratory measurement of an \Alfven wave collision, the
fundamental interaction underlying \Alfvenic turbulence.  By
interacting a relatively large-amplitude, low-frequency \Alfven wave
with a counterpropagating, smaller-amplitude, higher-frequency \Alfven
wave, the experiment accomplishes the secular nonlinear transfer of energy  to a
propagating daughter \Alfven wave.  The predicted properties of the
nonlinearly generated daughter \Alfven wave are outlined, providing
a suite of  tests that can be used to confirm the successful
measurement of the nonlinear interaction between counterpropagating
\Alfven waves in the laboratory.
\end{abstract}

\maketitle 

\section{Introduction}
Turbulence is a ubiquitous phenomenon in space and astrophysical
plasmas, driving a cascade of energy from large to small scales and
strongly influencing the plasma heating resulting from the dissipation
of the turbulence. Turbulence is believed to play a key role in a wide
range of space and astrophysical plasma environments, influencing the
heating of the solar corona and acceleration of the solar
wind,\cite{McIntosh:2011} the dynamics of the interstellar
medium,\cite{Armstrong:1981,Armstrong:1995,Gaensler:2011} the
regulation of star formation,\cite{Rickett:1990} the transport of heat
in galaxy clusters,\cite{Peterson:2006} and the transport of mass and
energy into the Earth's magnetosphere.\cite{Sundkvist:2005}

At the large length scales and low frequencies characteristic of the
turbulence in astrophysical systems, the turbulent motions are
governed by the physics of \Alfven waves.\cite{Alfven:1942,Belcher:1971} In the idealized  framework of 
incompressible magnetohydrodynamics (MHD), it has been shown that the
turbulent cascade of energy from large to small scales is driven by
the nonlinear interaction between counterpropagating \Alfven waves,
\cite{Kraichnan:1965} the key concept underpinning modern theories of
Alfv\'enic turbulence.\cite{Sridhar:1994,Goldreich:1995,Boldyrev:2006}
This key concept remains at the foundation of theoretical
investigations of astrophysical turbulence, yet its applicability in
the moderately to weakly collisional conditions relevant to
astrophysical plasmas has never been observationally or experimentally
verified. Verification is necessary to establish a firm basis for the
use of simplified fluid models, such as incompressible MHD, to examine
the dynamics and consequences of turbulence in the weakly collisional
conditions of diffuse astrophysical plasmas. In addition, the detailed
nature of this nonlinear interaction forms the crucial distinction
between the two leading theories for strong MHD
turbulence.\cite{Goldreich:1995,Boldyrev:2006}

Several reasons make it unlikely that the nonlinear interaction
between counterpropagating \Alfven waves can ever be verified using
observations of turbulence in space or astrophysical environments.
Remote astrophysical observations cannot achieve sufficient spatial
resolution to identify individual \Alfvenic fluctuations.
\emph{In situ} spacecraft measurements in the near-Earth
solar wind are capable of sufficient resolution in time to identify the
\Alfvenic nature of fluctuations, but  yield information at only a 
single spatial point (or a few spatial points for multi-spacecraft
missions), insufficient to constrain the interaction between two
counterpropagating \Alfvenic fluctuations. Finally,  in all of these systems, 
the broad spectrum of turbulent modes confounds attempts to identify
the transfer of energy from two nonlinearly interacting \Alfven waves
to a third wave.  Only experimental measurements in the laboratory can
achieve the controlled conditions and high spatial resolution
necessary, but producing plasmas at the relevant magnetohydrodynamic
scales requires a large experimental plasma volume.  The unique
capabilities of the Large Plasma Device (LAPD) at UCLA
\cite{Gekelman:1991}, designed to study fundamental space plasma physics
processes, make possible the first laboratory measurement of the
nonlinear wave-wave interaction underlying \Alfvenic turbulence.\cite{Howes:2012b}

Howes and Nielson,\cite{Howes:2013a} hereafter Paper I, presents the
detailed derivation of an asymptotic analytical solution for the
evolution of an \Alfven wave collision in the weakly nonlinear limit
using the incompressible MHD equations. The fundamental insight into
the nonlinear dynamics gleaned from this analytical solution lays the
theoretical foundation for the experimental design described
here. Nielson, Howes, and Dorland,\cite{Nielson:2013a} hereafter Paper
II, presents a numerical verification of this analytical solution and
demonstrates that the nonlinear interaction in a weakly collisional
astrophysical plasma remains well described by the incompressible MHD
solution.  In this paper, we describe in detail the theoretical
considerations used to design the first experiment to measure the
nonlinear interaction between counterpropagating
\Alfven waves.\cite{Howes:2012b} A companion work by Drake \emph{et
al.},\cite{Drake:2013} hereafter Paper IV, describes in detail the
experimental setup and procedure and the analysis used to identify
conclusively in the laboratory the nonlinear product of two
counterpropagating \Alfven waves.  Finally, in Howes \emph{et
al.},\cite{Howes:2013c} hereafter Paper V, the results of gyrokinetic
numerical simulations of \Alfven wavepacket collisions are used to
illustrate a magnetic shear interpretation of the nonlinear energy
transfer occurring due to the counterpropagating \Alfven wave collision
in the experiment designed here.

\section{Theory of \Alfven Wave Collisions}
\label{sec:theory}
Modern theories of anisotropic \Alfvenic plasma turbulence, relevant
to a wide range of space and astrophysical environments, have been
developed largely based on several key concepts derived from the
equations of incompressible MHD. Here we review some fundamental
properties of \Alfven wave collisions that provide the theoretical
foundation for the design of an experiment to measure the nonlinear
interaction between counterpropagating \Alfven waves in the
laboratory.

\subsection{Basic Properties}
\label{sec:basic}
As shown in Paper I, the equations of incompressible MHD can be
written in a symmetrized Els\"asser form,\citep{Elsasser:1950}
\begin{equation}
\frac{\partial \V{z}^{\pm}}{\partial t} 
\mp \V{v}_A \cdot \nabla \V{z}^{\pm} 
=-  \V{z}^{\mp}\cdot \nabla \V{z}^{\pm} -\nabla P/\rho_0,
\label{eq:elsasserpm}
\end{equation}
\begin{equation}
\nabla^2 P/\rho_0 =  - \nabla \cdot \left(  \V{z}^{-}\cdot \nabla \V{z}^{+}\right)
\label{eq:press}
\end{equation}
where the magnetic field is decomposed into equilibrium and
fluctuating parts $\V{B}=\V{B}_0+ \delta \V{B} $, $\V{v}_A
=\V{B}_0/\sqrt{4 \pi\rho_0}$ is the \Alfven velocity due to the
equilibrium field $\V{B}_0=B_0 \zhat$, $P$ is total pressure (thermal
plus magnetic), $\rho_0$ is mass density, and $\V{z}^{\pm}(x,y,z,t) =
\V{u} \pm \delta \V{B}/\sqrt{4 \pi \rho_0}$ are the Els\"asser 
fields given by the sum and difference of the velocity fluctuation
$\V{u}$ and the magnetic field fluctuation $\delta \V{B}$ expressed in
velocity units. 

The symmetrized Els\"asser form of the incompressible MHD equations
lends itself to a particularly simple physical interpretation.  
The Els\"asser field $\V{z}^{+}$ represents either the
\Alfven or pseudo-\Alfven wave traveling down the equilibrium magnetic
field (in the anti-parallel direction), while $\V{z}^{-}$ represents
either one of these waves propagating up the equilibrium magnetic
field (in the parallel direction).  The second term on the left-hand
side of \eqref{eq:elsasserpm} is the
\emph{linear term} representing the propagation of the Els\"asser
fields along the mean magnetic field at the \Alfven speed, the first
term on the right-hand side is the \emph{nonlinear term} representing
the interaction between counterpropagating waves, and the second term
on the right-hand side is a nonlinear term that ensures
incompressibility through \eqref{eq:press}.

Paper I describes in detail the numerous linear and nonlinear
properties of these equations, and here we highlight the principal
properties that significantly influence the evolution of turbulence in
an incompressible MHD plasma. First, it is shown that the \Alfven
waves dominate the nonlinear dynamics of the
turbulence\cite{Schekochihin:2009} in the anisotropic limit, $k_\perp
\gg k_\parallel$, a limit that naturally develops in magnetized plasma
turbulence.\cite{Robinson:1971,Belcher:1971,
Zweben:1979,Montgomery:1981,Shebalin:1983,Cho:2000,Maron:2001,
Cho:2004,Cho:2009,Sahraoui:2010b,Narita:2011,TenBarge:2012a} Second,
the nonlinear interaction between two \Alfven waves occurs only when
those \Alfven waves are propagating in opposite directions along the
equilibrium magnetic field.\cite{Iroshnikov:1963,Kraichnan:1965}
Third, the nonlinear interaction between two counterpropagating plane
\Alfven waves with wavevectors $\V{k}^+$ and  $\V{k}^-$ is proportional 
to $\zhat \cdot (\V{k}_{\perp}^- \times \V{k}_\perp^+)$, where
perpendicular is defined with respect to the direction of the
equilibrium magnetic field, $\V{B}_0=B_0 \zhat$. This implies that a
nonzero nonlinear interaction between counterpropagating \Alfven waves
requires that the two waves must have perpendicular components that
are not colinear.  Together, these properties dictate that \emph{the
fundamental building block of turbulence in an incompressible MHD
plasma is the nonlinear interaction between perpendicularly polarized,
counterpropagating
\Alfven waves.}

\subsection{Imbalanced Wave Amplitudes}
\label{sec:imbalance}

Another property of \Alfven wave collisions that can be exploited in
the design of an experiment to measure this nonlinear interaction
involves using two counterpropagating \Alfven waves
of different amplitudes.  This property is apparent if we rewrite the
equation for the evolution of the Els\"asser field $\V{z}^{+}$ in the
following form,
\begin{equation} 
\left[ \frac{\partial }{\partial t}
 - \V{v}_A \cdot \nabla \right] \V{z}^{+} =
-(\V{z}^{-}\cdot \nabla) \V{z}^{+} - \nabla P/\rho_0.
\label{eq:elsasser_z+}
\end{equation}
Noting that \eqref{eq:press} shows that the pressure depends linearly
on $\V{z}^{+}$, the entire equation \eqref{eq:elsasser_z+} is linear
in $\V{z}^{+}$. Ignoring the linear dependence on $\V{z}^{+}$ that
appears in every term of the equation, the terms on the right-hand side of
\eqref{eq:elsasser_z+}, which are responsible for the
nonlinear evolution of the $\V{z}^{+}$ field, depend on the amplitude
of the $\V{z}^{-}$ field, but not on the amplitude of the $\V{z}^{+}$
field.  For example, for significantly different amplitudes
$|\V{z}^{+}| \ll |\V{z}^{-}|$, the small-amplitude \Alfven wave
corresponding to $\V{z}^{+}$ will be strongly distorted nonlinearly by
the large-amplitude $\V{z}^{-}$ \Alfven wave, but the large-amplitude
$\V{z}^{-}$ \Alfven wave will only be weakly distorted by the
small-amplitude $\V{z}^{+}$ \Alfven wave. This interaction between
waves of unequal amplitudes is related to the topic of imbalanced
plasma turbulence (for the case of MHD turbulence, also referred to
as turbulence with nonzero cross helicity).
\cite{Lithwick:2007,Beresnyak:2008,Chandran:2008,Perez:2009,Podesta:2010b}

Therefore, to measure the nonlinear interaction between
counterpropagating \Alfven waves, it is sufficient to produce only one
large-amplitude \emph{distorting} wave, and to measure the nonlinear
distortion of a small-amplitude \emph{probe} wave.  Note that the
nonlinear distortion of the small amplitude probe wave, when its
waveform is decomposed into Fourier modes, is represented by the
transfer of energy into Fourier modes not present in the undistorted
probe wave.

\subsection{Weak MHD Turbulence}
\label{sec:weakturb}
For sufficiently small wave amplitudes, the nonlinear terms on the
right-hand side of \eqref{eq:elsasserpm} are small compared to the
linear term, and one obtains a state of \emph{weak MHD turbulence}.
\cite{Sridhar:1994}  In the weak turbulence paradigm, two
counterpropagating \Alfven waves may interact nonlinearly to transfer
energy to a third mode---this is the fundamental interaction that
underlies the cascade of energy from large to small scales in a
turbulent plasma. The small amplitude of the nonlinear term makes it
possible to derive an analytical solution for the nonlinear evolution
of weak MHD turbulence using perturbation
theory.\cite{Galtier:2000,Howes:2013a} It is important to note that
the linear term in \eqref{eq:elsasserpm} has no counterpart in
incompressible hydrodynamics, so it is not possible to define a state
of weak incompressible hydrodynamic turbulence.  Incompressible
hydrodynamic turbulence is always strong, precluding the application
of perturbation theory to obtain an analytical solution. This is an
important, and often underappreciated, fundamental distinction between
the turbulence in incompressible hydrodynamic systems and that in
incompressible MHD systems.

Applying perturbation theory to the case of two counterpropagating
plane \Alfven waves with wavevectors $\V{k}_1$ and $\V{k}_2$, the
lowest-order nonlinear interaction is the three-wave interaction with
a third mode with wavevector $\V{k}_3$. When averaged over an integral
number of  wave periods, this three-wave interaction satisfies the
resonance conditions
\begin{equation}
\V{k}_1+ \V{k}_2 = \V{k}_3 \quad \mbox{ and } \quad \omega_1 + \omega_2 = \omega_3.
\label{eq:constraints}
\end{equation}
These constraints are equivalent to the conservation of momentum and
conservation of energy.\cite{Sridhar:1994,Galtier:2000}

During the late 1990's, significant controversy arose regarding the
nature of weak MHD turbulence, focused on the question of whether the
lowest-order, nonzero nonlinear interactions are three-wave or
four-wave interactions. The history of this controversy is thoroughly
outlined in Paper I, but here we briefly discuss a few relevant
details because our aim is to design an experiment that leads to a
nonzero three-wave interaction.

Consider the nonlinear interaction between two counterpropagating
plane \Alfven waves with wavevectors $\V{k}_1$ and $\V{k}_2$. The
linear dispersion relation gives the \Alfven wave frequency in terms
of the component of the wavevector parallel to the equilibrium
magnetic field, $\omega=|k_\parallel| v_A$. Here we adopt the
convention that the frequency is always positive, $\omega>0$, so that
the sign of $k_\parallel$ indicates the direction of propagation of
the wave along the equilibrium magnetic field. As discussed earlier, a
nonzero nonlinearity requires counterpropagating
\Alfven waves, implying that $k_{\parallel 1}$ and $k_{\parallel
2}$ have opposite signs. It has been pointed out that the only
nontrivial solution to both constraints in \eqref{eq:constraints}
therefore has either $k_{\parallel 1}=0$ or $k_{\parallel
2}=0$.\cite{Shebalin:1983} The consequence of this finding is that
there is no cascade of energy to higher parallel wavenumber in weak
MHD turbulence.  Further, since propagating plane \Alfven waves require
nonzero $k_{\parallel}$, this led early researchers to posit that the
three-wave interaction in weak MHD turbulence---the lowest order in
the perturbative expansion---was zero, and that the four-wave
interaction represented the lowest-order, nonzero nonlinear
interaction.\cite{Sridhar:1994} The hypothesis that three-wave
interactions were empty was later demonstrated to be false because, in
a turbulent astrophysical system, the tendency for adjacent magnetic
field lines to wander away from each other manifests itself as an
effective $k_{\parallel}=0$ component to the
turbulence.\cite{Ng:1996,Galtier:2000,Lithwick:2003} Therefore, the
leading order of turbulent energy transfer in a weakly turbulent MHD
plasma obeys the constraints imposed in \eqref{eq:constraints},
and no parallel cascade of energy occurs.

\subsection{Weakly Nonlinear \Alfven Wave Collisions}
\label{sec:anal}

In the design of a laboratory experiment to measure an \Alfven wave
collision, the importance of whether the lowest-order nonlinear
interaction between counterpropagating \Alfven waves is a three-wave
or four-wave interaction cannot be overstated. Consider the
interaction between two counterpropagating \Alfven waves, each with
amplitude $\delta B$, in a magnetized plasma with an equilibrium
magnetic field of magnitude $B_0$. The
amplitude\cite{Sridhar:1994,Ng:1996} of a mode generated by three-wave
interactions is proportional to $(\delta B/B_0)^2$, whereas the
amplitude of a mode generated by four-wave interactions is
proportional to $(\delta B/B_0)^3$.  Since achievable wave amplitudes
in the laboratory are typically small compared to the equilibrium
magnetic field, $\delta B / B_0 \ll 1$, the signal due to a four-wave
interaction will be significantly smaller than that arising from a
three-wave interaction.

In order to develop valuable physical intuition about the nature of
\Alfven wave collisions, in Paper I we derived an asymptotic
analytical solution for the nonlinear interaction between
counterpropagating \Alfven waves in the weakly nonlinear limit. Here
we briefly outline the properties of that solution and use the
resulting insight to identify the necessary aspects of the
experimental design to achieve a nonzero three-wave nonlinear
interaction in the laboratory. A more detailed qualitative description
of this analytical solution is provided in \S IV.A of Paper~I.

The asymptotic solution employs the incompressible MHD equations to
solve for the evolution of the nonlinear interaction between two
counterpropagating plane \Alfven waves, with wavevectors $\V{k}_1^+ =
k_{\perp 0} \xhat - k_{\parallel 0}
\zhat$ and $\V{k}_1^- = k_{\perp 0} \yhat + k_{\parallel 0} \zhat$, where the 
equilibrium magnetic field is $\V{B}_0=B_0 \zhat$. The plasma inhabits
a periodic domain of size $L_\parallel \times L_\perp^2$ that is
elongated along the direction of the equilibrium magnetic field such
that $L_\parallel \gg L_\perp$.  The wavenumber components of the
initial \Alfven waves are at the domain scale, $k_{\perp 0}  \equiv 2 \pi
/L_\perp$ and $k_{\parallel 0} \equiv 2 \pi /L_\parallel$, so that problem
models the nonlinear dynamics in the \emph{anisotropic limit}, $k_{\perp } 
\gg k_\parallel$. In this limit, the \Alfvenic dynamics decouples from 
the dynamics of the pseudo-\Alfven waves, and the nonlinear evolution
of the \Alfvenic fluctuations is rigorously described by the
Els\"asser potential equations.\cite{Schekochihin:2009} The
characteristics of the nonlinear solution outlined below are described
using a shorthand notation for the spatial Fourier modes,
$(k_x/k_{\perp 0} , k_y/k_{\perp 0} , k_z/k_{\parallel 0})$, such that the initial
\Alfven waves are denoted by $\V{k}_1^+ =(1,0,-1)$ and $\V{k}_1^-
=(0,1,1)$.

The lowest-order nonlinear solution arises from a three-wave
interaction in which the primary \Alfven waves $\V{k}_1^+ $ and
$\V{k}_1^-$ interact to transfer energy to a secondary mode
$\V{k}_2^{(0)}=(1,1,0)$. This mode is a strictly magnetic fluctuation
with no variation along the equilibrium magnetic field, $k_\parallel=0$, but
with an oscillation frequency $2 \omega_0$. Since this mode does not
satisfy the linear \Alfven wave dispersion relation, it is an
inherently nonlinear fluctuation. In addition, the nonlinear
interaction between $\V{k}_1^+ $ and $\V{k}_1^-$ also generates an
electromagnetic standing wave caused by the superposition of two
counterpropagating linear \Alfven waves with wavevectors
$\V{k}_2^{(-2)}=(-1,1,2)$ and $\V{k}_2^{(+2)}=(1,-1,-2)$, each with
frequency $2 \omega_0$. The amplitude of all three of these Fourier
modes, each generated by nonlinear three-wave interactions, oscillates
in time as shown by equations (5)--(8) in Paper II, so there is no
secular transfer of energy for the lowest-order nonlinear solution.

Since the interacting primary \Alfven waves have no $k_\parallel=0$
component, there is no net energy transfer for the lowest-order
nonlinear solution (resulting from three-wave interactions),
consistent with the findings of Ng and Bhattacharjee.\cite{Ng:1996}
Physically, since the equilibrium magnetic field defined in our
periodic domain does not wander---the relevant situation for a
laboratory plasma experiment confined by a uniform, axial magnetic
field---the three-wave interaction leads to zero net transfer of
energy at any time equal to a half-integral number of primary wave
periods, $t=n\pi/\omega_0$ for $n=1,2,3,\ldots$. This case contrasts
with the argument for turbulent astrophysical plasmas, described in
\secref{sec:weakturb}, in which field-line wander leads to an
effective $k_\parallel=0$ component to the turbulence, enabling a net
transfer of energy due to three-wave interactions.

The secular transfer of energy first arises in this problem due to
four-wave interactions appearing at the next order of the asymptotic
solution in Paper I. When the secondary mode $\V{k}_2^{(0)}$ has
nonzero amplitude, the two primary \Alfven waves $\V{k}_1^+$ and $\V{k}_1^-$
nonlinearly interact with it to transfer energy to two tertiary
\Alfven waves $\V{k}_3^+ = (2,1,-1)$ and $\V{k}_3^-
=(1,2,1)$, respectively.  The amplitudes of these tertiary \Alfven
waves increase linearly with time, indicating a secular transfer of
energy.  The net effect of combining these two three-wave
interactions---the interaction $\V{k}_1^+ +\V{k}_1^-=\V{k}_2^{(0)}$
followed by $\V{k}_1^\pm +
\V{k}_2^{(0)}=\V{k}_3^\pm $---is equivalent to a four-wave interaction that
causes the secular transfer of energy from the primary \Alfven waves
at low wavenumber with $|\V{k}_1^\pm | =\sqrt{k_{\perp 0}^2 +
k_{\parallel 0}^2}$ to the tertiary \Alfven waves at higher wavenumber
with $|\V{k}_3^\pm | =\sqrt{5k_{\perp 0}^2 + k_{\parallel 0}^2}$.  In
agreement with the heuristic model of weak
turbulence,\cite{Sridhar:1994,Montgomery:1995,Goldreich:1997,Galtier:2000,Lithwick:2003}
this nonlinear energy transfer is strictly perpendicular, manifesting
a perpendicular cascade of energy to higher $k_\perp$, but no parallel
cascade of energy to higher $k_\parallel$.

\subsection{Implications for Experimental Design}
Exploiting the physical insight derived from the theoretical
considerations outlined above in this section, we discuss here the
characteristics of the experiment necessary to achieve a measurable
nonlinear interaction between counterpropagating \Alfven waves in the
laboratory.

The experiment\cite{Howes:2012b} was performed using the Large Plasma
Device (LAPD),\cite{Gekelman:1991} a basic plasma science user
facility operated at the University of California, Los Angeles. This
experimental apparatus generates a plasma in a cylindrical column of
16.5~m length and 40~cm diameter confined by a background axial
magnetic field.  The basic setup is to employ two separate antennas,
located at opposite ends of the plasma chamber, to launch \Alfven
waves in opposite directions along the equilibrium axial magnetic
field. The wave magnetic fields of the two linearly polarized \Alfven
waves are oriented perpendicular to each other to maximize the
nonlinear interaction, satisfying the desired properties outlined in
\secref{sec:basic}.

The two antennas used for the experiment, a Loop
antenna\citep{Auerbach:2011} and an Arbitrary Spatial Waveform (ASW)
antenna,\citep{Thuecks:2009,Kletzing:2010} have different capabilities
that can be exploited to optimize the resulting nonlinear signal in
the experiment. The Loop antenna can generate a relatively
large-amplitude \Alfven wave (up to $\delta B/B_0 \sim 0.01$, although
in this experiment we use $\delta B/B_0 \sim 0.002$), but the spatial
waveform in the plane perpendicular to the axial magnetic field
consists of a complicated combination of Fourier modes, making the
identification of any nonlinear distortion of this wave difficult.
The ASW antenna, on the other hand, can generate a spatial waveform
dominated by a single plane-wave Fourier mode, simplifying the task of
measuring the nonlinear energy transfer to a different Fourier
component, but is limited to much smaller
\Alfven wave amplitudes ($\delta B/B_0 \sim 0.00002$). Therefore, we
choose to perform the experiment with counterpropagating \Alfven
waves of unequal amplitude, using a large-amplitude \emph{distorting}
\Alfven wave  launched by the Loop antenna to cause  nonlinear evolution of
a small-amplitude \emph{probe} \Alfven wave launched by the ASW
antenna, taking advantage of the properties of the nonlinear
interaction discussed in \secref{sec:imbalance}. 

The instrumental limitation to small \Alfven wave amplitudes compared
to the equilibrium magnetic field, $\delta B /B_0 \ll 1$, leads to a
system with dynamics that satisfy the weakly nonlinear limit.  In
addition, the plasma volume in the experiment is highly elongated
along the equilibrium magnetic field, leading to nonlinear dynamics
within the anisotropic limit, $k_\perp \gg k_\parallel$.  Finally, the
strong equilibrium magnetic field in the experiment is usually chosen
to be as straight and uniform as possible to yield good plasma
confinement, so the equilibrium magnetic field lines do not wander, in
contrast to the astrophysical case. Fortunately, these experimental
limitations lead to a system that is directly comparable to the
asymptotic analytical solution outlined in \secref{sec:anal}, so we
can use this solution to guide the experimental design.

Our goal is to devise an experiment that will generate a propagating
\emph{daughter} \Alfven wave due to a secular transfer of  energy 
from the primary interacting \Alfven waves.  It is important that this
nonlinearly generated daughter \Alfven wave have a sufficient
amplitude to be measurable in the laboratory, even in the weakly
nonlinear limit dictated by instrumental limitations.  Therefore, it
is desirable that this energy transfer be governed by a three-wave
interaction, which produces a significantly larger amplitude signal
than a four-wave interaction.  However, for the symmetric \Alfven wave
collision problem defined in \secref{sec:anal}, in which the two
primary \Alfven waves have equal and opposite values of $k_\parallel$,
the three-wave interaction between these primary \Alfven waves does
not yield a secular transfer of energy. The four-wave interaction, on
the other hand, does lead to a net secular transfer of energy to a
propagating daughter \Alfven wave. This four-wave interaction consists
of the sequence of three-wave interactions $
\V{k}_1^+ +\V{k}_1^-=\V{k}_2^{(0)}$ followed by $\V{k}_1^\pm +
\V{k}_2^{(0)}=\V{k}_3^\pm $, where the nonlinearly generated magnetic
fluctuation with $k_\parallel=0$ plays a key role in mediating the energy
transfer. However, the amplitude of the daughter \Alfven wave generated
nonlinearly by this four-wave interaction is unlikely to
have a sufficient signal-to-noise ratio to be measurable in the
laboratory.

Upon examination of the two sequential three-wave interactions
constituting the four-wave interaction in the symmetric problem, the
second of these interactions, $\V{k}_1^\pm +
\V{k}_2^{(0)}=\V{k}_3^\pm$, has the property we desire: the nonlinear
transfer of energy from one propagating \Alfven wave $\V{k}_1^\pm$ to
another propagating \Alfven wave $\V{k}_3^\pm$ with higher
perpendicular wavenumber.  Therefore, we require that the waveform of
one of the propagating \Alfven waves, in this case the large-amplitude
distorting \Alfven wave, include a $k_\parallel=0$ component. This will
satisfy the requirement, discussed in \secref{sec:weakturb}, that one
of the interacting modes in the three-wave interaction has $k_\parallel=0$.
This property can be achieved in the laboratory by breaking the
symmetry between the parallel wavenumbers of the two primary
counterpropagating \Alfven waves.

In our approach, the Loop antenna launches a large-amplitude
distorting \Alfven wave $\V{z}^-$ up the axial magnetic field.  This
distorting \Alfven wave has a longer parallel wavelength,
$\lambda_\parallel^-= 2 \pi/ k_\parallel^-$, than the parallel
wavelength, $\lambda_\parallel^+$, of the counterpropagating,
small-amplitude probe \Alfven wave $\V{z}^+$ launched by the ASW
antenna, such that $\lambda_\parallel^->
\lambda_\parallel^+$. 
 Furthermore, the distorting antenna parallel wavelength is longer
 than twice the physical distance $L$ over which the two
 counterpropagating waves interact, $\lambda_\parallel^-> 2 L$. As
 demonstrated quantitatively in \secref{sec:design}, for this case,
 the section of the waveform of the distorting \Alfven wave that
 interacts with the probe \Alfven wave will, in general, have a
 nonzero $k_\parallel=0$ component.  Therefore, the distorting \Alfven
 wave can mediate the secular transfer of energy from the primary
 probe \Alfven wave, through a nonlinear three-wave interaction, to a
 propagating daughter \Alfven wave. The resulting nonlinear evolution
 is consistent with the findings of Ng and Bhattacharjee\cite{Ng:1996}
 that nonlinear energy transfer due to three-wave interactions is
 nonzero when the interacting wavepackets have a $k_\parallel=0$
 component. In \secref{sec:packet} below, we discuss in more detail
 the physical significance of an \Alfven wavepacket with a
 $k_\parallel=0$ component.

Note that, experimentally, the parallel wavelength of an \Alfven wave
launched by an antenna in the LAPD is controlled by the driving
frequency of the antenna. The argument above can be cast into a
complementary form using the temporal basis of period or frequency
instead of wavelength or wavenumber.  For \Alfven waves in the MHD
limit, which are non-dispersive, the linear dispersion relation
connecting these alternative descriptions is simply $\lambda_\parallel
=v_A T$.  The equivalent temporal description is that we use a
low-frequency distorting \Alfven wave to nonlinearly modify a
counterpropagating higher-frequency probe
\Alfven wave. The time over which the two waves interact is less than
the period of the low-frequency distorting \Alfven wave, leading to a
nonzero net energy transfer mediated by a three-wave interaction. Note
that all of the \Alfven wave frequencies used in the experiment must
be somewhat below the ion cyclotron frequency, typically $\omega
\lesssim \Omega_i/2$, to prevent the cyclotron resonance from significantly
altering the \Alfven wave dynamics.\cite{Nielson:2010}

\subsection{Plane Waves versus Wavepackets}
\label{sec:packet}

The controversy over the existence of three-wave interactions in weak MHD
turbulence arose partly through the consideration of a problem that
was implicitly restricted to the interactions between plane \Alfven
waves, where each plane wave mode is described by its single wavevector,
$\V{k}$.  For a plane \Alfven wave to propagate, the parallel
component of this wavevector must be nonzero. The physical reason is
that magnetic tension is the restoring force that supports the
propagation of \Alfven waves, and only modes with $k_\parallel \ne 0$
bend the equilibrium magnetic field. If one has a single spatial
Fourier mode with $k_\parallel=0$, the magnetic field is not bent, so
there arises no magnetic tension to support \Alfven wave propagation.

\begin{figure}[top]
\resizebox{86mm}{!}{\includegraphics*[0.29in,5.6in][8.1in,9.9in]{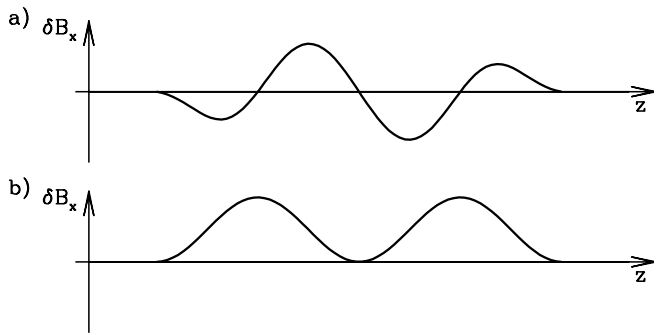}}
\caption{Schematic diagram of the waveforms of two \Alfven wavepackets. 
(a) A waveform that is symmetric about $\delta B_x=0$, and that therefore
has no $k_\parallel=0$ component. (b) An asymmetric waveform that
includes a significant $k_\parallel=0$ component.
\label{fig:packet}}
\end{figure}

But a more physically relevant case to consider than the interaction
between two plane \Alfven waves of infinite extent is the interaction
between two spatially localized \Alfven wavepackets.\cite{Ng:1996}
This more general case of wavepacket interactions leads to a
significantly different perspective on the role of $k_\parallel=0$
modes, as follows.  First, as shown in Paper I, we note that, in the
absence of any \Alfven wave energy propagating in the opposite
direction ($\V{z}^+=0$), a finite amplitude \Alfven wave $\V{z}^-$ of
arbitrary waveform is an exact\ nonlinear solution to
\eqref{eq:elsasserpm}. Any such \Alfven wavepacket $\V{z}^-$ will
propagate undistorted up the equilibrium magnetic field at the
\Alfven speed, $v_A$. Next, for an incompressible MHD plasma with an
equilibrium magnetic field $\V{B}_0=B_0 \zhat$, we consider the two
different waveforms of \Alfven wavepackets presented in
\figref{fig:packet}.  The waveform in panel (a) is symmetric about
$\delta B_x=0$, so a Fourier decomposition of this wavepacket has no
$k_\parallel=0$ component. The waveform in panel (b), on the other
hand, is not symmetric about $\delta B_x=0$, and therefore a Fourier
decomposition  yields a significant $k_\parallel=0$ component to
this \Alfven wavepacket. Yet, in the absence of any counterpropagating
\Alfven wavepackets, $\V{z}^+=0$, both of these \Alfven wavepackets will
propagate similarly up the equilibrium magnetic field at the \Alfven
speed without any nonlinear distortion, even for finite wave amplitude.

When a counterpropagating \Alfven wavepacket $\V{z}^+$ with no
$k_\parallel=0$ component collides with one of the waveforms in
\figref{fig:packet}, however, the nonlinear evolution differs 
significantly for these two cases.  For the case involving the
symmetric waveform in panel (a), there will be no three-wave
interaction since neither wavepacket has a $k_\parallel=0$
component. The lowest-order nonlinear interaction in this case is the
four-wave interaction.\cite{Sridhar:1994} For the case involving the
asymmetric waveform in panel (b), however, the $k_\parallel=0$
component of the $\V{z}^-$ wavepacket leads to a three-wave
interaction that nonlinearly distorts the counterpropagating $\V{z}^+$
\Alfven wavepacket.

In a turbulent astrophysical plasma, of course, there is no reason
that the many \Alfven wavepackets that constitute the turbulence need
have symmetric waveforms. In general, one would expect the waveform of
any given \Alfven wavepacket not to be symmetric, so it is therefore
expected that three-wave interactions will dominate in the case of
weak turbulence in an astrophysical plasma.\cite{Ng:1996} In Paper V,
we show that the $k_\parallel=0$ component of an \Alfven wavepacket is
equivalent to a magnetic shear, establishing the connection between
magnetic field line wander and plasma turbulence.

The key point to take away from this discussion is that an asymmetric
\Alfven waveform will generate a three-wave interaction because its
plane wave decomposition contains a finite $k_\parallel=0$ component.
In the laboratory, we aim to design an experiment that will generate
such an asymmetric \Alfven waveform, and will therefore lead to
nonlinear energy transfer through a three-wave interaction. In the
next section, we demonstrate quantitatively that, for the particular
experimental setup chosen, the waveform of the large-amplitude
distorting \Alfven wave includes the desired $k_\parallel=0$
component.

\section{Experimental Design}
\label{sec:design}
The aim of this paper is to explain the theoretical considerations
underlying the design of an experiment to measure the product of the
nonlinear interaction between perpendicularly polarized,
counterpropagating \Alfven waves in the laboratory.  The experiment is
designed to measure the interaction between a large-amplitude,
low-frequency \emph{distorting} \Alfven wave and a counterpropagating
smaller-amplitude, higher-frequency \emph{probe} \Alfven wave. In this
section, we demonstrate that the nonlinear interaction in this
experimental setup will generate a propagating
\Alfven wave from the lowest-order, three-wave interaction, and we
 predict the properties of this nonlinear \emph{daughter} \Alfven
 wave.

\begin{figure}[top]
\resizebox{86mm}{!}{\includegraphics{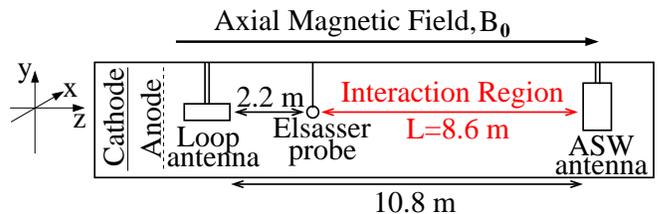}}
\caption{Schematic of the \Alfven wave turbulence experiment on the LAPD. 
The Loop antenna generates a large-amplitude distorting \Alfven wave,
with the wave magnetic field dominantly polarized in the
$x$-direction, traveling up the equilibrium axial magnetic field,
$\V{B}_0=B_0 \zhat$. The ASW antenna generates a small-amplitude probe \Alfven wave
polarized in the $y$-direction traveling down the axial  magnetic field.
\label{fig:setup}}
\end{figure}

A simplified schematic of the LAPD experiment is shown in
\figref{fig:setup}. The Loop antenna\cite{Auerbach:2011} generates 
a relatively large-amplitude, distorting \Alfven wave $\V{z}^-$, with its wave
magnetic field dominantly polarized in the $x$-direction, traveling up
the equilibrium axial magnetic field, $\V{B}_0=B_0 \zhat$. The
Arbitrary Spatial Waveform (ASW)
antenna\cite{Thuecks:2009,Kletzing:2010} generates a small-amplitude,
probe \Alfven wave $\V{z}^+$, polarized in the $y$-direction, traveling down the
axial magnetic field.  The magnetic and electric field signals of the waves are measured by an Els\"asser probe\cite{Drake:2011}
located near the Loop antenna. The probe
\Alfven wave is nonlinearly distorted by the counterpropagating 
Loop \Alfven wave over the length of the \emph{interaction region}
between the ASW antenna and the Els\"asser probe, $L=8.6$~m, as
depicted in \figref{fig:setup}. 

If the parallel wavelength of the distorting \Alfven wave
$\lambda_\parallel^- > 2L$, then the probe \Alfven wave will interact
with only a fraction of a wavelength of the distorting wave before it
is measured by the Els\"asser probe. In this case, the net effect of
the three-wave interaction is nonzero because the distorting wave
effectively includes a $k_\parallel=0$ component. The experiment described
here, and in our companion work Paper IV,\cite{Drake:2013} is aimed at
measuring the nonlinear outcome of that interaction in the laboratory.

\subsection{The Waveform of the Distorting Wave over the Interaction Region}

\begin{figure}[top]
\resizebox{86mm}{!}{\includegraphics*[0.5in,5.8in][7.8in,9.9in]{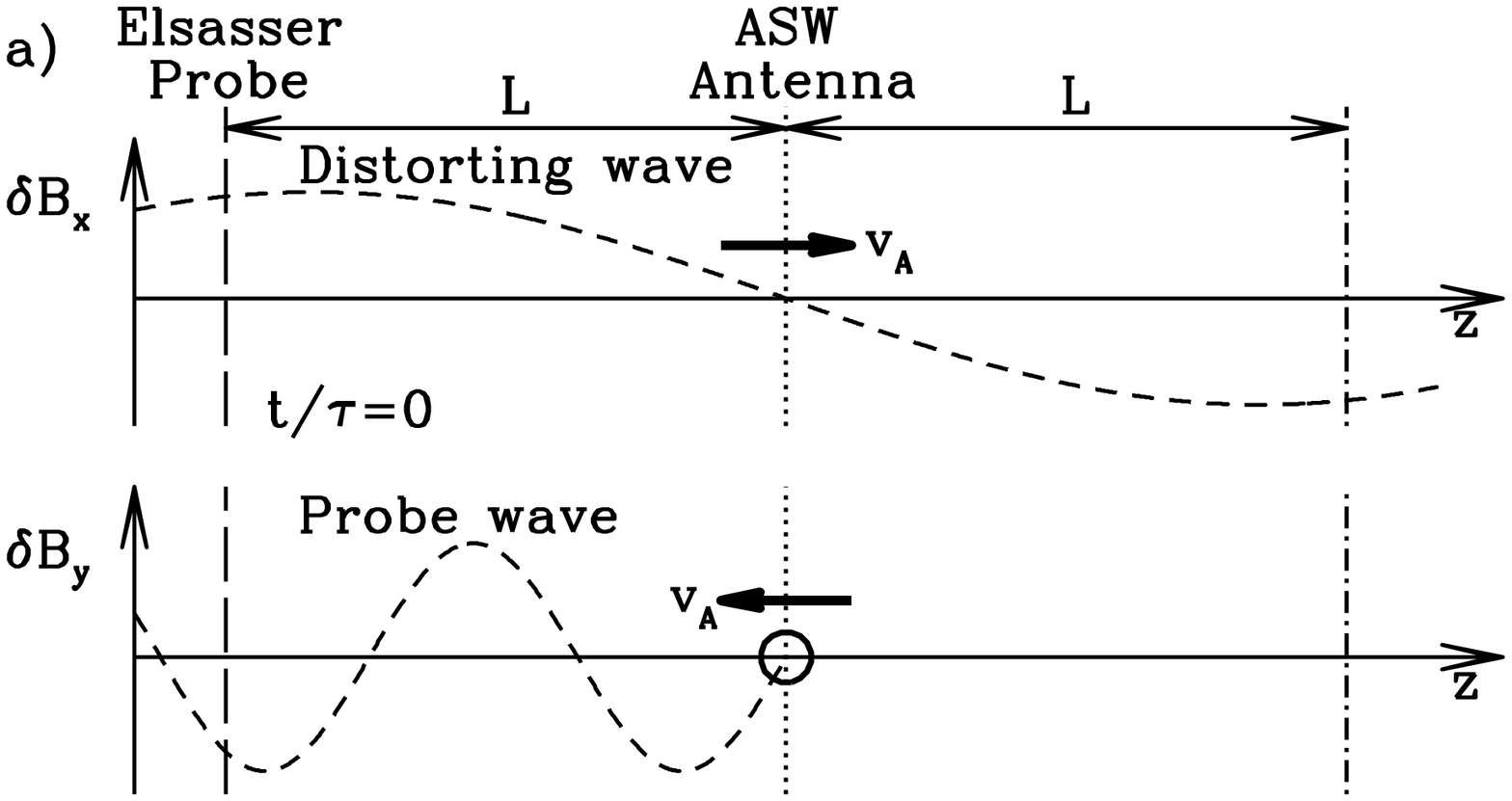}}
\resizebox{86mm}{!}{\includegraphics*[0.5in,5.8in][7.8in,9.9in]{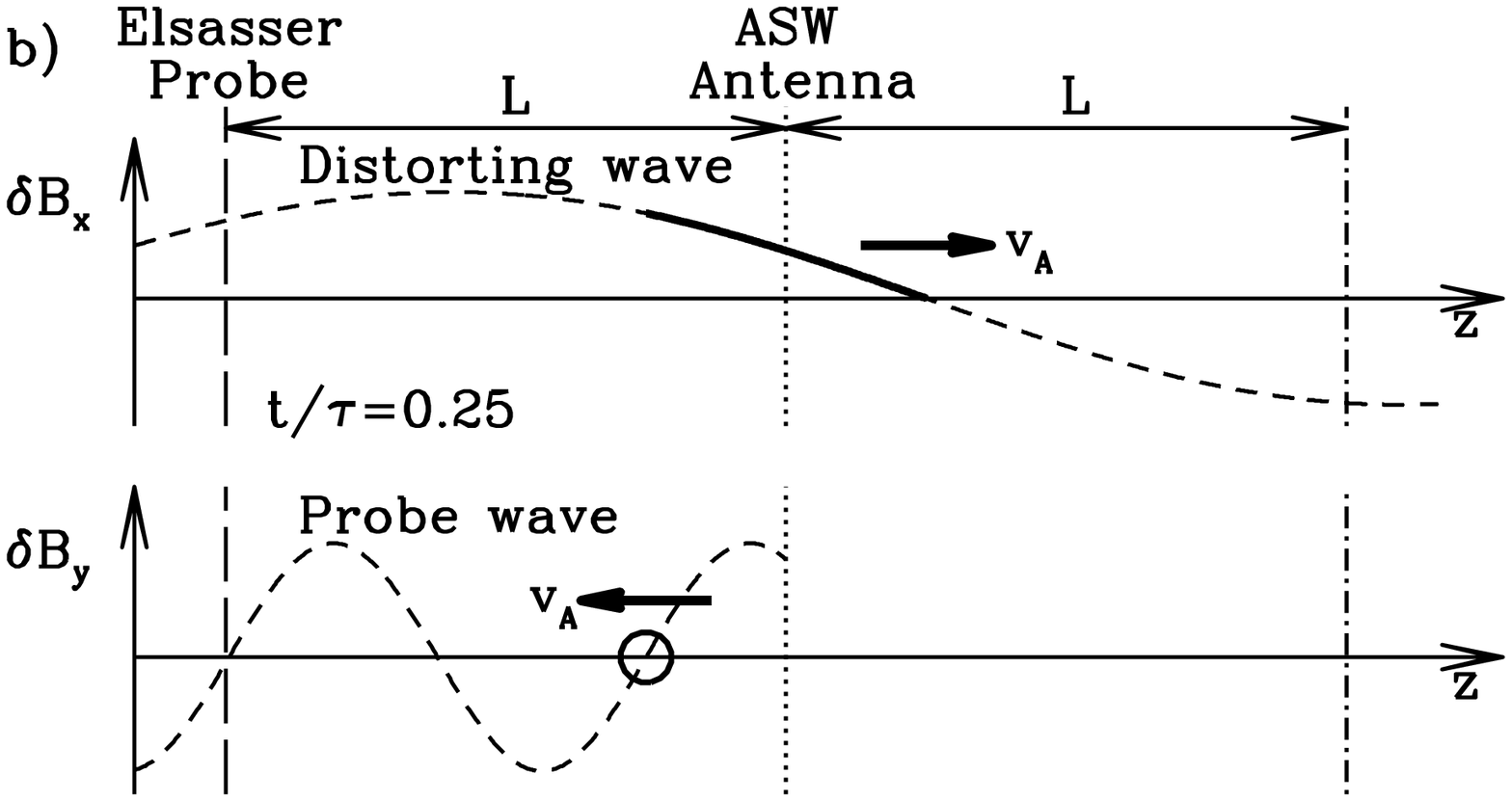}}
\resizebox{86mm}{!}{\includegraphics*[0.5in,5.8in][7.8in,9.9in]{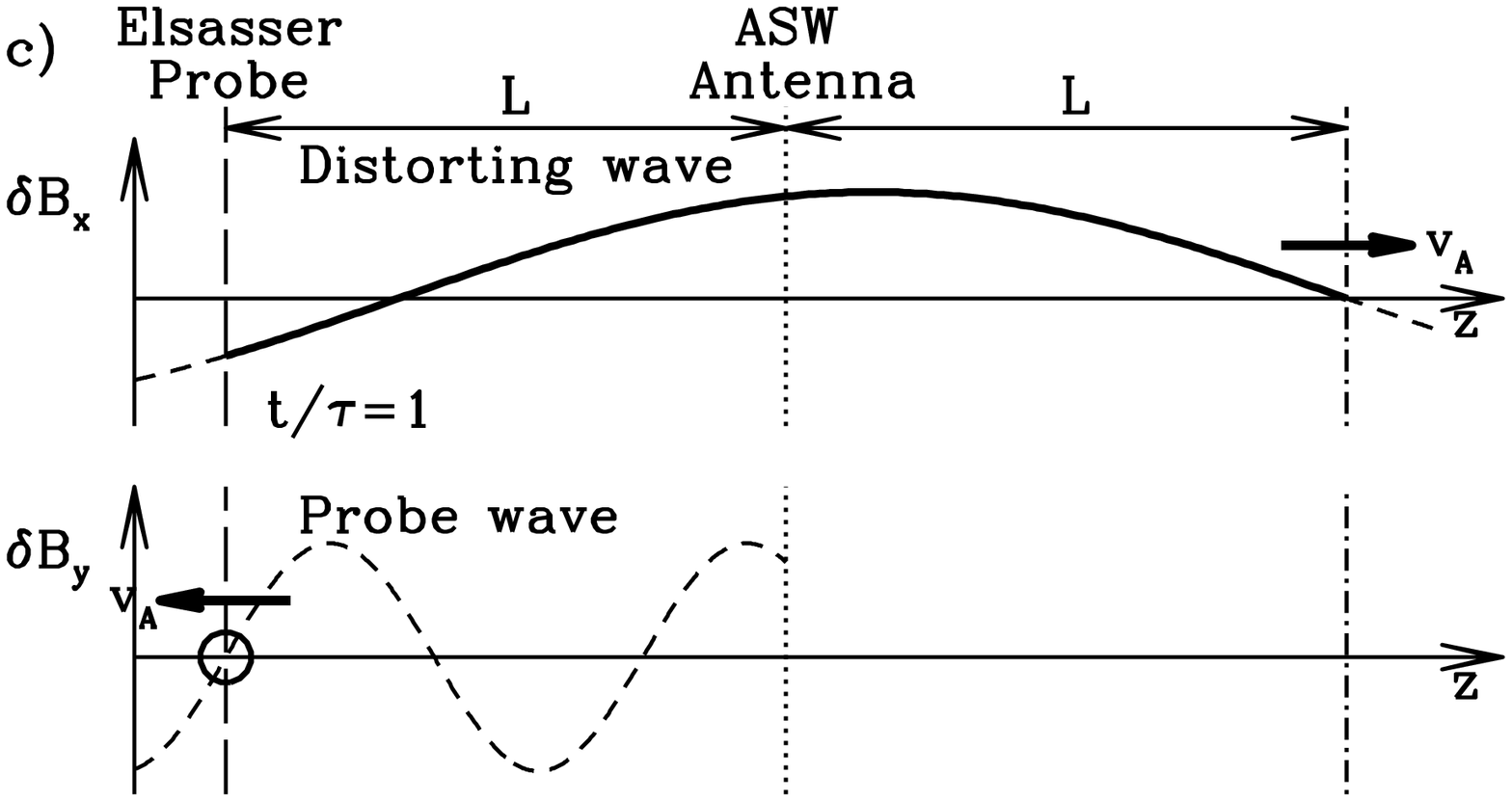}}
\caption{Each panel depicts the $\delta B_x$ component of the 
rightward propagating distorting Loop \Alfven wave (upper) and the
$\delta B_y$ component of the leftward propagating probe ASW \Alfven
wave (lower). The length of the distorting \Alfven wave that has
interacted with one point on the probe \Alfven wave (open circle) is
depicted by the thick solid line in the upper plot of each panel.
\label{fig:interaction}}
\end{figure}

For a sufficiently long parallel wavelength of the distorting \Alfven
wave $\V{z}^-$ such that $\lambda_\parallel^- > 2L$, each point on the
counterpropagating probe \Alfven wave $\V{z}^+$ interacts with only
a fraction of the waveform before it is measured by the Els\"asser probe. To
demonstrate this quantitatively, it is instructive to illustrate this
interaction as a function of time, as depicted in
\figref{fig:interaction}.  
All of the parameters of the LAPD experiment are provided in Paper
IV,\cite{Drake:2013} so here we merely quote the values of the
parameters relevant to this discussion. The parallel wavelength of the
distorting Loop \Alfven wave is $\lambda_\parallel^- = 29.1$~m and of
the probe ASW \Alfven wave is $\lambda_\parallel^+ = 6.4$~m. The
interaction region spans $L=8.6$~m along the equilibrium magnetic
field between the ASW antenna and the Els\"asser probe, as shown in
\figref{fig:setup}.

In \figref{fig:interaction}, the distorting Loop \Alfven wave travels
to the right at the \Alfven velocity $v_A$ in the upper plot of each
panel (the Loop antenna is physically located just to the left of the
ordinate axis). Similarly, the lower plot of each panel shows the
probe  \Alfven wave traveling to the left at the \Alfven velocity
$v_A$, launched from the ASW antenna (center, vertical dotted line)
towards the Elsasser probe (left, vertical long-dashed line).  Note
that the length $L$ on the left side, between the Elsasser probe and
the ASW antenna, corresponds to the interaction region in the
experiment.  The length $L$ to the right of the ASW antenna is
included to show the part of the distorting \Alfven wave that has
passed the position of the ASW antenna (physically, this part of the
wave is disrupted by the presence of the ASW antenna).

Let us consider the length of the distorting
\Alfven wave that has interacted with a particular point on the probe
\Alfven wave (marked by an open circle in the lower plot of each 
panel in \figref{fig:interaction}). We plot the length of the
distorting \Alfven wave that has interacted (thick solid line) as a
function of time normalized by $\tau=v_A/L$, the probe wave travel
time from the ASW antenna (center, vertical dotted line) to the
Els\"asser probe (left, vertical long-dashed line). In panel (a), at
$t/\tau=0$, the point on the probe wave is just leaving the ASW
antenna (lower), and has not interacted with any of the distorting
wave (upper). In panel (b), at $t/\tau=0.25$, the particular point on
the probe wave (open circle) has moved one quarter of the distance
from the ASW antenna to the Els\"asser probe (lower). The section of
the distorting wave that has interacted with the particular point of
the probe wave (upper) is indicated by the thick solid line. Note that
this section of thick solid line is twice the distance that the probe
\Alfven wave has traveled since the relative speed between the two
counterpropagating waves is $2 v_A$. In panel (c), at $t/\tau=1$, the
point on the probe wave has reached the Els\"asser probe where it is
measured (lower).  That point on the probe wave has interacted with
the portion of the distorting wave (upper) that spans a length $2L$
(thick solid line).

The plots in \figref{fig:interaction} illustrate that the nonlinear
interaction is the result of a length $2L$ of the distorting wave
interacting with each point on the probe wave. The key concept here is
that, as long as $\lambda_\parallel^- > 2L$, the probe wave interacts
with only a fraction of the wavelength of the distorting wave.  Even
though an integral number of wavelengths of the distorting wave has no
$k_\parallel =0$ component, the waveform over a length that is a
non-integral number of wavelengths  contains a
$k_\parallel =0$ component.  The resulting waveform of the distorting
\Alfven wave that is responsible for nonlinearly modifying the probe
wave is indicated by the thick solid line in the upper plot of panel
(c).  In the next section, we show that this part of the distorting
\Alfven wave contains an effective $k_\parallel =0$ component.

\subsection{Fourier Components of the Distorting Wave}
\label{sec:fourier}

To understand how the experimental design we have chosen is able to
generate nonlinearly a propagating daughter \Alfven from the
lowest-order, three-wave interaction, we must consider the section of
the distorting Loop \Alfven wave, as depicted by the thick solid line
in the upper plot of panel (c) in \figref{fig:interaction}, that is
responsible for causing a nonlinear distortion of the probe \Alfven
wave. We can Fourier transform to decompose that waveform into its
plane wave components, and consider the nonlinear interaction of each
of those components of the distorting wave with the probe wave.

\begin{figure}[top]
\resizebox{86mm}{!}{\includegraphics{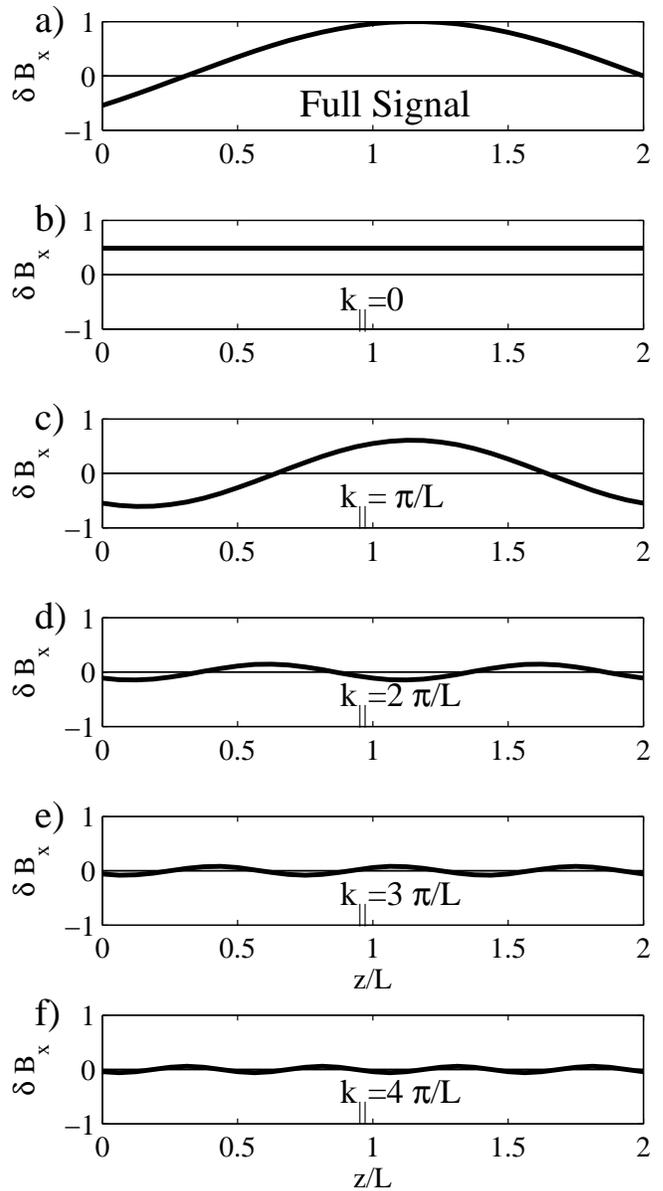}}
\caption{Fourier decomposition of the distorting \Alfven wave signal over a 
length $2L$, showing that the waveform contains a significant
$k_\parallel=0$ component.
\label{fig:decompose}}
\end{figure}

The Fourier decomposition of the distorting \Alfven wave is shown in
\figref{fig:decompose}. In panel (a) is shown the full waveform of the
distorting \Alfven wave, the same waveform as that given by the thick
solid line in the upper part of panel (c) in \figref{fig:interaction}.
The parallel length of the signal is $2L$, so the components of the
Fourier transform have parallel wavenumbers $k_\parallel = 0, \pi/L,
2\pi/L, 3\pi/L, \ldots$, and the lowest five Fourier modes are plotted
in panels (b)--(f).

\figref{fig:decompose} makes clear the key point that, although the distorting Loop
\Alfven wave is simply a single sinusoidal component with wavelength
$\lambda_\parallel^- = 29.1$~m, when sampled over only a fraction of
its wavelength, it can have a significant $k_\parallel=0$ component
that interacts with the counterpropagating probe ASW \Alfven wave.
This effective $k_\parallel=0$ component, shown in panel (b), is in
fact the energetically dominant component for the particular case
examined in \figref{fig:decompose}. The $k_\parallel=\pi/L$ mode also
contains significant energy, but the higher-order components with
$k_\parallel>\pi/L$ contain very little energy.  Therefore, for the
particular case examined here, the dominant components of the
distorting \Alfven wave that interact with the probe \Alfven wave are
the two lowest wavenumber components of the Fourier series,
$k_\parallel=0$ and $k_\parallel=\pi/L$.

It is worthwhile pointing out that the results of the Fourier
decomposition shown in \figref{fig:decompose} depend on the what
section (of length $2L$) of the distorting \Alfven wave is sampled. As
a window of length $2L$ is moved over the Loop antenna waveform, which
has a parallel wavelength $\lambda_\parallel^- = 29.1$~m, the
coefficient of the $k_\parallel=0$ Fourier component oscillates about
zero.  This coefficient reaches an extremum when the window is
centered on an extremum of the distorting Loop antenna wave (so that
the windowed signal has even parity about the center), and passes
through zero when the window is centered on a zero crossing of the
distorting Loop antenna wave (so that the windowed signal has odd
parity). As we shall see, this property implies that the measured
amplitude of the nonlinearly generated daughter \Alfven wave will
increase and decrease in time.

\subsection{Nonlinear Interactions}
\label{sec:nl}

The Fourier decomposition of the distorting \Alfven wave above can be
used to split the total nonlinear interaction between the probe
\Alfven wave and the distorting \Alfven wave into the sum of the
interactions between the probe \Alfven wave and each Fourier component
of the distorting \Alfven wave. Let us consider each of these
interactions in turn, beginning with the energetically dominant
components of the distorting \Alfven wave with $k_\parallel=0$ and
$k_\parallel=\pi/L$, as shown in \figref{fig:decompose}.

Based on the theoretical considerations discussed in
\secref{sec:theory}, the $k_\parallel=0$ component of the
distorting \Alfven wave will mediate a three-wave interaction to
transfer energy secularly from the probe \Alfven wave to a daughter
\Alfven wave.  This nonlinear interaction is the desired aim of the
experimental design, and we determine the properties of the resulting
daughter \Alfven wave in \secref{sec:properties}.

The other energetically dominant component of the distorting \Alfven
wave, with $k_\parallel=\pi/L$, leads, on the other hand, to no net
nonlinear energy transfer due to the three-wave interaction occurring
as the probe wave traverses the interaction region. A secular energy
transfer can occur due to a four-wave interaction, as explained in
\secref{sec:anal}, but such an effect is probably not measurable in
the laboratory due to the significantly smaller amplitude of any mode
generated by a four-wave interaction.

As it turns out, all of the higher wavenumber components (with
$k_\parallel>\pi/L$) of the distorting \Alfven wave will likewise lead
to zero net energy transfer due to three-wave interactions.  This is
most easily understood in terms of the magnetic shear associated with
each of these modes, as described in Paper V. In general, the
three-wave interaction will not produce any net nonlinear energy
transfer if the integral of the magnetic fluctuation $\delta B_x$ over
the length $z=2L$ is zero, as is the case for all of these modes,
easily verified by inspection of \figref{fig:decompose}. Paper V
demonstrates this point using numerical simulations of collisions
between counterpropagating \Alfven wave packets.

In conclusion, only the $k_\parallel=0$ component of the distorting
\Alfven wave, as experienced by the counterpropagating probe \Alfven
wave over the length of the interaction region, will lead to a secular
transfer of energy from the probe \Alfven wave to a daughter \Alfven
wave. This is consistent with findings of Ng and
Bhattacharjee\cite{Ng:1996} that a nonzero $k_\parallel=0$ component
of an \Alfven wavepacket is required to yield nonzero nonlinear energy
transfer via a resonant three-wave interaction.

\subsection{Predicted Properties of the Daughter Wave}
\label{sec:properties}
Let us now predict the properties of the nonlinearly generated
daughter \Alfven wave arising from experimental design described here.
As we have shown in \secref{sec:nl}, only the $k_\parallel=0$
component of the distorting \Alfven wave leads to a measurable
transfer of energy from the probe \Alfven wave to a propagating
daughter \Alfven through a resonant three-wave interaction. 

First, let us define the wavevectors of the plane-wave Fourier modes
associated with each of the interacting, counterpropagating \Alfven
wave components. The probe \Alfven wave is launched by the ASW antenna
down the magnetic field in the $-\zhat$ direction with the
perpendicular waveform shown in panel (b) of Figure~2 in Paper IV. The
perpendicular spatial variation of this waveform is almost entirely in
the $\xhat$ direction. Therefore, we denote the wavevector of the
probe ASW \Alfven wave $\V{z}^+$ by $\V{k}_1= k_{\perp 1} \xhat + k_{\parallel
1} \zhat$, where $k_{\parallel 1}<0$ indicates that the wave
propagates anti-parallel to the axial magnetic field shown in
\figref{fig:setup}. 

Although the distorting \Alfven wave $\V{z}^-$ launched by the Loop
antenna has a more complicated perpendicular waveform, as shown in
panel (b) of Figure~3 in Paper IV, we focus on the variation of the
$\delta B_x$ component of the wave magnetic field.  In
\eqref{eq:elsasserpm}, the  vector form of the nonlinear 
term, $\V{z}^- \cdot \nabla \V{z}^+$, indicates that it is the $\delta
B_x$ component of the distorting \Alfven wave $\V{z}^-$ that will
interact nonlinearly with the probe \Alfven wave $\V{z}^+$; this
follows because the perpendicular component of the wavevector
$\V{k}_1$ of the probe \Alfven wave is in the $\xhat$-direction. As
shown in panel (b) of Figure~7 in Paper IV, the perpendicular Fourier
transform of the $\delta B_x$ component of the distorting Loop \Alfven
wave has very little power with $k_x \ne 0$, so the wavevector is
dominantly in the $y$-direction. Only the component with $k_{\parallel
2}=0$ of the distorting Loop \Alfven wave leads to a net nonlinear
energy transfer through the three-wave interaction, so the wavevector of
the relevant Fourier mode for the distorting \Alfven wave is given by
$\V{k}_2= k_{\perp 2} \yhat$. For simplicity in the discussion below,
when we refer to distorting
\Alfven wave, we specifically consider \emph{only} this Fourier
component $\V{k}_2$ of that wave, since the other components lead to
zero net energy transfer via three-wave interactions.

The wavevector of the daughter \Alfven wave resulting from the
nonlinear interaction between the distorting and probe \Alfven waves
in the experiment is denoted by $\V{k}_3 =\V{k}_{\perp 3} +
k_{\parallel 3}\zhat$. The resonance conditions for the three-wave
interactions given by
\eqref{eq:constraints} enable us to predict  $\V{k}_{\perp 3}$ 
and $k_{\parallel 3}$ in terms of $k_{\perp 1}$, $k_{\parallel 1}$,
$k_{\perp 2}$, and $k_{\parallel 2}$.

Before applying the resonance conditions \eqref{eq:constraints}, we
need to determine the frequency $\omega_2$ associated with the Fourier
mode $\V{k}_2$ of the distorting \Alfven wave.  As illustrated by
panel (b) of \figref{fig:decompose}, the perpendicular magnetic field
fluctuation $\delta B_x$ associated with the $k_{\parallel 2}=0$ component
of the distorting
\Alfven wave is constant as experienced by any single point on the
counterpropagating probe \Alfven wave. Therefore, this component will
have zero frequency, $\omega_2=0$, from the perspective of the
counterpropagating wave. The probe \Alfven wave launched by the ASW
antenna is a linear \Alfven wave,  so its frequency is given
by the linear dispersion relation, $\omega_1=|k_{\parallel 1}| v_A$, where we 
reiterate that $k_{\parallel 1}<0$ to denote that the probe \Alfven wave
travels down the axial magnetic field in the experiment.

Now that we have specified the wavevectors and frequencies of the
distorting and probe \Alfven waves that lead to energy transfer via a
resonant three-wave interaction, we can compute the properties of the
nonlinearly generated daughter \Alfven wave.  The resonance conditions
\eqref{eq:constraints} simplify to the following two equations,
\begin{equation}
\V{k}_{\perp 1}+ \V{k}_{\perp 2} = \V{k}_{\perp 3} 
\label{eq:kperp}
\end{equation}
\begin{equation}
k_{\parallel 1} = k_{\parallel 3} 
\label{eq:kpar}
\end{equation}
We now discuss the implications of these properties for the
measurement of the daughter \Alfven wave in the laboratory.

The first distinguishing feature of the daughter \Alfven wave is that
its perpendicular wavevector is the vector sum of the perpendicular
wavevectors of distorting and probe \Alfven waves, $ \V{k}_{\perp 3}=
\V{k}_{\perp 1}+ \V{k}_{\perp 2}$.  Fourier transformation of the 
measured fluctuations over the perpendicular plane enables a clean
separation of the daughter \Alfven wave signal from its parent \Alfven
waves, as shown in panel (a) of Figure~7 in Paper IV.

A second technique for isolating the daughter wave signal is to
exploit the constraint from Maxwell's equations that the magnetic
field fluctuations associated with \Alfven waves must be divergence
free, $\nabla \cdot \delta \V{B}=0$. In the MHD limit relevant to the
proposed experiment, $k_\perp \rho_s \ll 1$, where $\rho_s$ is the ion sound Larmor radius, the \Alfven wave 
eigenfunction has a negligible parallel magnetic field fluctuation,
$\delta B_\parallel=0$. Therefore, the divergence-free condition for a
particular plane wave mode reduces to $\V{k}_\perp \cdot \delta
\V{B}_\perp=0$.  This means that the magnetic field fluctuation associated 
with a plane \Alfven wave is orthogonal to the perpendicular component
of the wavevector. This property suggests a simple way to separate the
signal of the daughter \Alfven wave from the probe \Alfven wave.  The
perpendicular wavevector of the probe \Alfven wave is $\V{k}_{\perp
1}= k_{\perp 1} \xhat $, so this divergence-free property dictates
that the probe \Alfven wave has $\delta B_{x 1}=0$. The daughter
\Alfven wave, on the other hand, has $\V{k}_{\perp 3}= k_{\perp 1}
\xhat + k_{\perp 2} \yhat$, so its magnetic field fluctuation is
polarized in the direction $\zhat \times \hat{\V{k}}_{\perp 3} =
(k_{\perp 1}/k_{\perp 3}) \yhat -(k_{\perp 2}/k_{\perp 3}) \xhat$,
where $k_{\perp 3}=\sqrt{k_{\perp 1}^2+ k_{\perp 2}^2}$. The relation
between the components of $\delta \V{B}_3$ is $\delta
B_{x3}=-(k_{\perp 2}/k_{\perp 1}) \delta B_{y3}$. Therefore, by
measuring the $\delta B_{x}$ component of the wave magnetic fields in
the experiment, the daughter \Alfven wave can be measured without
interference from the signal of the probe \Alfven wave.

A third property of the daughter \Alfven wave follows from the
parallel component of its wavevector, $ k_{\parallel 3} =k_{\parallel
1} $. Since the sign of $k_\parallel$ indicates the propagation
direction, this implies that the daughter \Alfven wave propagates in
the same direction as the probe \Alfven wave.  This also implies that
there is no parallel cascade of energy, consistent with the
expectations from weak turbulence
theory\cite{Shebalin:1983,Sridhar:1994,Ng:1996,Goldreich:1997} and the
asymptotic analytical solution derived in Paper I.

Following from $ k_{\parallel 3} =k_{\parallel 1} $ and the linear
dispersion relation for \Alfven waves, $\omega = |k_\parallel| v_A$, a
fourth distinguishing feature of the daughter \Alfven wave is that it
has the same frequency as the probe \Alfven wave,
$\omega_3=\omega_1$. Since the experimental design outlined in this
paper employs a much lower frequency distorting \Alfven wave launched
by the Loop antenna (considering here the full signal of this wave,
not just the $k_{\parallel 2}=0$ Fourier component), $\omega_2 < \omega_1$, this
enables the daughter \Alfven wave signal to be separated from the
distorting \Alfven wave by filtering in frequency.

The asymptotic analytical solution derived in Paper I suggests
additional distinguishing features of the daughter \Alfven wave. A
fifth property is that any daughter \Alfven wave receiving a secular
transfer of energy from a parent \Alfven wave should have a $\pi/2$
phase shift with respect to the parent wave, as indicated by the
solution for the secularly increasing mode in eq.~(40) of Paper
I. This prediction is supported by the numerical simulations presented
in Paper V.

A sixth property of the daughter \Alfven wave, suggested by the
analytical solution, is the predicted amplitude of the signal. Although
the analytical solution derived in Paper I is computed for a symmetric
case of perpendicularly polarized, counterpropagating \Alfven waves 
with the same magnitude of $k_\perp$ and equal and opposite values of
$k_\parallel$, the solution nonetheless should provide a reasonable
order-of-magnitude or better estimate of the amplitude of the signal
generated by nonlinear three-wave interactions. The coefficient of the
magnetic field perturbation due to the lowest-order nonlinear
solution, given by eq.~(36) in Paper I, is
\begin{equation}
 \frac{|\V{B}_{\perp 2}|}{B_0} \sim  \frac{  z_+ z_-}{16 v_A^2} 
\frac{  k_\perp }{k_\parallel}
\end{equation}
where $z_\pm/v_A$ represents the normalized amplitude of the Elsasser
fields $\V{z}^\pm$ associated with the primary counterpropagating
\Alfven waves. Using the linear eigenfunction for \Alfven waves,
$\V{u}_\perp/v_A = \pm \delta \V{B}_\perp /B_0$, we can relate the
Elsasser field amplitude to the perturbed magnetic field amplitude,
$|\V{z}^\pm|/v_A = 2 |\V{B}_\perp|/B_0$. Therefore, the relation for
the amplitude of the daughter wave $\delta B_{\perp 3}$ can be
simplified to
\begin{equation}
\frac{\delta B_{\perp 3}}{B_0}  \sim \frac{1 }{4}\frac{\delta B_{\perp 1}}{B_0}
\frac{\delta B_{\perp 2}}{B_0}
\frac{  k_{\perp 1} }{k_{\parallel 1}},
\label{eq:amp}
\end{equation}
where the $k_{\perp 1}$ and $k_{\parallel 1}$ are the wavevector components of
the probe \Alfven wave, $\delta B_{\perp 1}$ is the amplitude of the
probe \Alfven wave, and $\delta B_{\perp 2}$ corresponds to the
amplitude of only the $k_{\parallel 2}=0$ component of the distorting
\Alfven wave. This amplitude prediction can be used to determine whether 
the measured daughter \Alfven wave signal is consistent with
theoretical expectations.

A final qualitative characteristic of the daughter \Alfven wave signal
arises because the coefficient of the $k_{\parallel 2}=0$ component of
the distorting \Alfven wave oscillates in time, as mentioned in the
final paragraph of \secref{sec:fourier}.  Since the amplitude of the
$k_{\parallel 2}=0$ component varies depending on the section of the
distorting \Alfven wave that interacts with the probe
\Alfven wave over the interaction region, the net nonlinear energy
transfer to the daughter \Alfven wave oscillates with the amplitude of
the $k_{\parallel 2}=0$ component.  This $k_{\parallel 2}=0$ amplitude
oscillates at the frequency of the distorting \Alfven wave, so we
expect that amplitude of the nonlinear daughter wave signal will
increase and decrease at distorting \Alfven wave frequency.

\subsection{Relation to Recent Laboratory Experiments}
\label{sec:relation}

It is worth highlighting here the contrasts between the present
experimental investigation of \Alfven wave collisions as the
fundamental building block of plasma turbulence and a recent
experimental study exploring the physics of the parametric decay
instability using the nonlinear interaction between counterpropagating
\Alfven waves by Dorfman and Carter.\cite{Dorfman:2013}

It has long been known that finite-amplitude \Alfven waves are
nonlinearly unstable to parametric instabilities driven by gradients
parallel to the magnetic field, including the
decay,\cite{Galeev:1963,Sagdeev:1969}
modulational,\cite{Lashmore-Davies:1976} and
beat\cite{Wong:1986,Hollweg:1994} instabilities. In particular, the
parametric decay instability involves the nonlinear decay of a
large-amplitude \Alfven wave into an ion acoustic wave propagating in
the same direction and an \Alfven wave propagating in the
opposite direction.\cite{Sagdeev:1969} Thus, the parametric decay
instability relies on a nonlinear mode coupling between \Alfven and ion
acoustic waves.  Dorfman and Carter\cite{Dorfman:2013} performed the
first laboratory measurements of a related Alfv\'en-acoustic mode
coupling in which two counterpropagating \Alfven waves interact
nonlinearly to generate resonantly an ion acoustic mode at the
resulting beat frequency.

Although the experimental setup on the Large Plasma Device (LAPD) at
UCLA for the Dorfman and Carter study involved the nonlinear
interaction between counterpropagating \Alfven waves---and may
therefore appear very similar to the present study---the two
investigations, in fact, probe entirely distinct nonlinear physical
mechanisms. Our study probes the nonlinearity associated with the
perpendicular gradients of the wave magnetic fields, generically
labeled the $\V{E} \times \V{B}$ nonlinearity in Paper
I,\cite{Howes:2013a} whereas the Dorfman and Carter study explores the
nonlinear ponderomotive force due to parallel gradients of the
magnetic field magnitude.  The $\V{E} \times \V{B}$ nonlinearity
operates in an incompressible plasma, but requires that the
counterpropagating \Alfven waves have wavevectors $\V{k}^+$ and
$\V{k}^-$ with nonzero perpendicular components that are not colinear,
$\V{k}_\perp^+ \times \V{k}_\perp^- \ne 0$. Simply put, the $\V{E}
\times \V{B}$ nonlinearity occurs between counterpropagating,
perpendicularly polarized \Alfven waves. In contrast, the physics of
the parametric decay instability explored by Dorfman and Carter
requires compressibility, but does not demand that either \Alfven wave
have perpendicular spatial variation, occurring even when
$\V{k}_\perp^+ =\V{k}_\perp^- = 0$.  The wave magnetic fields, on the
other hand, cannot be perpendicularly polarized, so that $\delta
\V{B}_\perp^+ \cdot \delta \V{B}_\perp^- \ne 0$.  Simply put, the
nonlinearity studied by Dorfman and Carter occurs between
counterpropagating \Alfven waves polarized in the same direction.  It
is clear that these two experiments have been designed specifically to
probe distinct nonlinear mechanisms.

A number of studies have suggested that the nonlinearities associated
with the decay instability or other parametric instabilities play a
role in the nonlinear evolution of turbulence in the solar wind or
other astrophysical
plasmas.\cite{Derby:1978,Goldstein:1978,Spangler:1982,Sakai:1983,
Spangler:1986,Terasawa:1986,Wong:1986,Hollweg:1994,Shevchenko:2003}
However, the majority of the literature on parametric instabilities
adopts the oft-assumed ``slab'' geometry, a one-dimensional treatment
in which the waves vary only parallel to the magnetic field, so
$k_\perp=0$ for all modes. This limitation eliminates the possibility
of the $\V{E}\times \V{B}$ nonlinearity, a nonlinearity that will
dominate under conditions in which the turbulent fluctuations are
highly elongated along the local mean magnetic field, $k_\perp \gg
k_\parallel$, a property strongly supported by recent
multi-spacecraft observations of turbulence in the solar
wind.\cite{Sahraoui:2010b,Narita:2011,Roberts:2013}  In the
introduction to Paper I,\cite{Howes:2013a} we propose the working
hypothesis that the $\V{E} \times \V{B}$ nonlinearity is the dominant
nonlinear mechanism underlying the anisotropic cascade of energy in
magnetized plasma turbulence, and there we present a detailed argument
in support of this hypothesis.  Briefly, the argument is that
nonlinearities associated with parallel gradients, such as the
parametric instabilities, contribute significantly to the nonlinear
evolution only for finite amplitude fluctuations, $\delta v/v_A \sim
1$; on the other hand, the $\V{E} \times \V{B}$ nonlinearity depends
on the perpendicular gradients, and can therefore contribute
significantly even if $\delta v/v_A \ll 1$, as long as the turbulent
fluctuations are significantly anisotropic, $k_\perp / k_\parallel \gg
1$. Furthermore, that a nonlinear mechanism involving compressibility
is not required by \Alfvenic plasma turbulence is supported by the
fact that both incompressible\cite{Cho:2000,Maron:2001} and
compressible\cite{Cho:2003} MHD turbulence simulations recover
quantitatively similar turbulent energy spectra for the \Alfvenic
fluctuations.


\section{Conclusion}
\label{sec:conc}
In this paper, we describe the theoretical background of \Alfven wave
collisions used to design an experiment to measure in the laboratory
for the first time the nonlinear interaction between
counterpropagating \Alfven waves.  This successful measurement has
established a firm basis for the application of theoretical ideas
developed in idealized models to turbulence in realistic space and
astrophysical plasma systems.\cite{Howes:2012b} The theoretical
considerations here and the details of the experimental procedure and
analysis in a companion work by Drake \emph{et al.},\cite{Drake:2013}
(Paper IV) provide the necessary background for this experimental effort.

The experiment is designed to measure the propagating daughter \Alfven
wave generated by the nonlinear interaction between perpendicularly
polarized, counterpropagating \Alfven waves.  This nonlinear
interaction is the fundamental building block of astrophysical plasma
turbulence. In order to generate a measurable signal in the
experiment, a relatively large-amplitude, low-frequency \Alfven wave
is employed to distort nonlinearly a counterpropagating,
smaller-amplitude, higher-frequency \Alfven wave. The parallel
wavelength of the low-frequency \Alfven wave exceeds twice the length
of the experimental interaction region, generating an effective
$k_\parallel=0$ component that is experienced by the
counterpropagating high-frequency
\Alfven wave.  This $k_\parallel=0$ component mediates a resonant 
three-wave interaction that transfers energy secularly from initial
high-frequency \Alfven wave to a nonlinearly generated daughter
\Alfven wave.

For the three-wave interaction in the experiment between a
high-frequency \Alfven wave with wavevector $\V{k}_1= k_{\perp 1}
\xhat + k_{\parallel 1} \zhat$ and a counterpropagating low-frequency 
\Alfven wave with a $k_{\parallel 2}=0$ component 
with wavevector $\V{k}_2= k_{\perp 2}
\yhat$, we expect nonlinear energy transfer to a propagating daughter \Alfven wave
with wavevector $\V{k}_3 =\V{k}_{\perp 3} + k_{\parallel 3}\zhat$.  We
predict the following properties of the daughter \Alfven wave based on
the theoretical considerations outlined in this paper:
\begin{enumerate}
\item Perpendicular wavevector: $ \V{k}_{\perp 3}=
\V{k}_{\perp 1}+ \V{k}_{\perp 2}$.
\item Propagation direction:  $ k_{\parallel 3} =k_{\parallel 1} $.
\item Frequency: $\omega_3=\omega_1$.
\item Magnetic Field: $\delta
B_{x3}=-(k_{\perp 2}/k_{\perp 1}) \delta B_{y3}$.
\item Phase: Daughter \Alfven wave phase-shifted by $\pi/2$ from the 
probe \Alfven wave.
\item Amplitude: \\$\delta B_{\perp 3}/B_0  \sim (1/4)(\delta B_{\perp 1}/B_0)
(\delta B_{\perp 2}/B_0)
(  k_{\perp 1} /k_{\parallel 1})$
\end{enumerate}
These theoretically predicted properties of the daughter \Alfven wave
can be used to demonstrate conclusively that our
experiment\cite{Howes:2012b} has indeed lead to the first successful
measurement of an \Alfven wave generated by the nonlinear interaction
between counterpropagating \Alfven waves, as detailed in Paper IV.

A critical point of this analysis is that any localized \Alfven
wavepacket with an asymmetric waveform carries a $k_\parallel=0$
component. This $k_\parallel=0$ component can lead to the secular
transfer of energy to \Alfven waves with higher perpendicular
wavenumber through resonant three-wave interactions. Since \Alfven
wavepackets in weakly turbulent astrophysical plasmas will not
generally have symmetric waveforms, three-wave interactions are
expected to dominate the nonlinear energy transfer.  The experimental
design described here has successfully demonstrated in the
laboratory\cite{Howes:2012b} the nonlinear energy transfer by resonant
three-wave interactions in a weakly turbulent plasma system in the MHD
regime. 

In a  forthcoming work (Paper V\cite{Howes:2013c}), nonlinear
gyrokinetic simulations of the interaction between localized, counterpropagating 
\Alfven wavepackets will be used to  illustrate this concept of 
energy transfer by resonant three-wave interactions when asymmetric
\Alfven waveforms collide. Furthermore, the  $k_\parallel=0$
component of an \Alfven wavepacket will be interpreted physically as a
finite magnetic shear, establishing a firm connection between the concepts
of magnetic field line wander and turbulence in astrophysical plasmas.

\begin{acknowledgments}
This work was supported by NSF PHY-10033446, NSF CAREER AGS-1054061,
NSF CAREER PHY-0547572, and NASA NNX10AC91G. This works describes the
design of an experiment conducted at the Basic Plasma Science
Facility, funded by the U.S. Department of Energy and the National
Science Foundation.
\end{acknowledgments}


%

\end{document}